\documentclass[twocolumn,a4paper,superscriptaddress,noeprint,prx]{revtex4-2}%
\usepackage{graphicx}
\usepackage{dcolumn}
\usepackage{bm}
\usepackage{xcolor}
\usepackage{breqn}
\usepackage{geometry}
\usepackage{soul}
\usepackage{algorithm}
\usepackage{algpseudocode}
\usepackage{enumitem} 
\newcommand{\stkout}[1]{\ifmmode\text{\sout{\ensuremath{#1}}}\else\sout{#1}\fi}

\begin{document}

\preprint{APS/123-QED}

\title{Topology Controls the Phase Separation Dynamics \\ of Many Component Fluid Mixtures}

\author{Michael Rennick}%
\affiliation{$School\ of\ Engineering,\ University\ of\ Edinburgh,\ Edinburgh\ EH9\ 3FD,\ United\ Kingdom$}
\author{Xitong Zhang}%
\affiliation{$School\ of\ Engineering,\ University\ of\ Edinburgh,\ Edinburgh\ EH9\ 3FD,\ United\ Kingdom$}
\author{Halim Kusumaatmaja}
\email{halim.kusumaatmaja@ed.ac.uk}
\affiliation{$School\ of\ Engineering,\ University\ of\ Edinburgh,\ Edinburgh\ EH9\ 3FD,\ United\ Kingdom$}

\date{\today}

\begin{abstract}
Fluid mixtures, ranging from the cellular cytoplasm to synthetic DNA nanostar systems, can spontaneously compartmentalize into many ($N$) coexisting liquid phases through liquid-liquid phase separation. While such systems exhibit a remarkable diversity of spatial organizations, the physical principles governing their non-equilibrium dynamics remain poorly understood. Here, combining simulations and analytical theory, we show that the coarsening dynamics of many component phase separation are fundamentally linked to mathematical coloring problems. 
For planar phase organization, relevant to synthetic droplet monolayers and simple biological structures, we identify distinct topological constraints for $N=2$, $N=3$, and $N=4$, with no further change for $N>4$, consistent with the four-color theorem. These constraints govern the coarsening dynamics, and, using chromatic graph theory, we derive a theoretical model for $N\geq3$ that quantitatively captures the diffusive-like coarsening. By contrast, classical theories based solely on Ostwald ripening underestimate the observed dynamics.
We further show that tuning interfacial tensions modifies the set of admissible phase arrangements, enabling highly heterogeneous coarsening dynamics across different phases. For unconfined systems with nonplanar phase organization, different coloring constraints apply, with no analogue of the four-color theorem, and coalescence suppression emerges only when the number of phases exceeds $N\gtrsim7$. More broadly, our work establishes coloring theory as a topological framework for understanding and predicting the dynamics of many component phase-separating fluids.
\end{abstract}

\maketitle

\section{Introduction}

Spontaneous phase separation of fluid mixtures offers a powerful mechanism for spatial organization in living and synthetic systems, enabling the formation of multiple coexisting phases with distinct compositions. For example, liquid-liquid phase separation in the cellular cytoplasm partitions hundreds of different proteins and nucleic acids into chemically distinct condensates that support a wide range of biological functions~\cite{phasesepcondensate,KilgoreScience,Banani2017,Choi,Berry_2018,Erkamp2023,Mehta2022}. The number of unique phases that spontaneously separate then depends on the interactions between the biomolecules \cite{zwicker,phasesepmany1,phasesepmany2,zcvb-9t4b}, resulting in fluid compartments with diverse morphologies and subcompartments~\cite{FERIC20161686,Chen2024,Wan2018,doyle2023,SANDERS2020306,Ye2025.05.14.654140,Wan2018,Fare2021,Zhao2024,Fei2017}. Synthetically, recent advances in DNA and RNA self-assembly have allowed fine-tuning of biomolecular interactions leading to complex mesostructures, including the possibility to realize many (up to nine so far) immiscible fluid phases~\cite{chaderjian2025diversedistinctdenselypacked,Fabrini2024,Wilken2023,GongAFM,Abraham_2024,BiffiPNAS}.

Despite some progress in understanding the thermodynamics of spontaneous phase separation of many fluid components~\cite{Jacobs2021,Rodrigo2024,teixeira2025metastablephaseseparationinformation,Krishna2021} and the influence of interfacial tensions on the equilibrium phase morphologies~\cite{phaseseptheo1,phaseseptheo2}, it is still unclear how cells or synthetic systems can leverage the number of coexisting phases to control both the spatial organization and coarsening dynamics of fluid compartments. This question is particularly important because phase separated structures emerge dynamically rather than being assembled directly at equilibrium, and coarsening dynamics determine which phase structures are ultimately realized on relevant timescales. Extensive works on the dynamics of fluid phase separation both in two and three-dimensions have largely focused on two immiscible components, and it is well understood that there can be several different dominant coarsening dynamics, including diffusive, viscous hydrodynamics, and inertial hydrodynamics~\cite{phasesepscaling,phasesepscalingsource,Wagner_2001,WagnerYeomans}. However, it remains an open question how the growing number of interactions between fluid components reshape the coarsening dynamics when more than two phases coexist.

\begin{figure}[!t]
	\centering
	\includegraphics[width=\columnwidth]{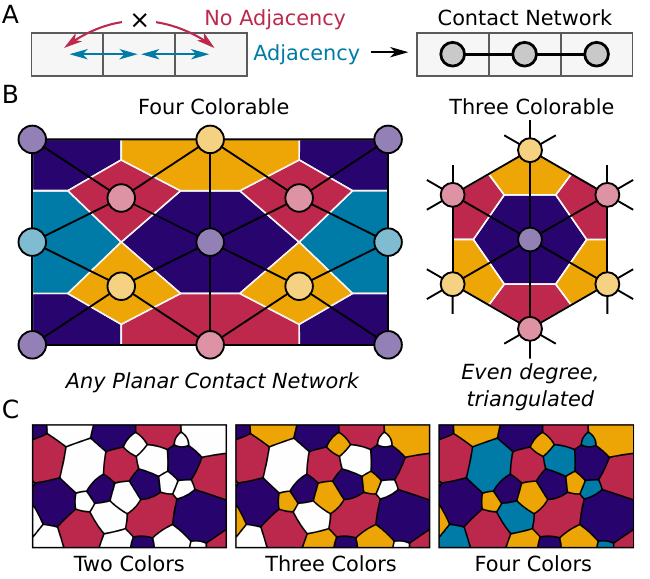}
	\caption{Schematic representation of contact networks and associated coloring theorems. (A) Contact networks represent the spatial adjacencies of different fluid regions in the domain. We place a node of the network within each region and connect the network with every adjacency that fulfils the following condition: two nodes are adjacent if a continuous path can be drawn between them without passing through any fluid region associated with other nodes. (B) Illustration of the four-color theorem, which guarantees that any planar contact network can be colored with four colors without adjacency of identical colors~\cite{ROBERTSON19972}. A contact network is planar if it can be projected onto a plane without edge crossings. Triangulated planar contact networks where each vertex has an even number of neighbors are a special class of three-colorable graphs~\cite{STEINBERG1993211}. (C) Example of a domain that requires at least four colors. White regions cannot be filled without adjacency of like colors.}
	\label{fig:fourcol}
\end{figure}

Here we investigate phase separation dynamics with up to eight fluid phases, employing lattice Boltzmann simulations that incorporate full hydrodynamics coupling. We analyse the resulting dynamics through the evolving contact network between fluid domains, illustrated in Fig.~\ref{fig:fourcol}(A). Because coalescence quickly eliminates interfaces between identical phases, adjacent regions in the contact network must contain different fluids.
The permissible domain configurations are therefore intimately connected to mathematical coloring problems, which ask how graphs can be colored without adjacent vertices sharing the same color. A familiar example is the four-color theorem~\cite{ROBERTSON19972}, which guarantees that every planar graph is four-colorable, as shown in Fig.~\ref{fig:fourcol}(B,C). 

For planar phase organization, relevant to synthetic droplet monolayers~\cite{chaderjian2025diversedistinctdenselypacked,Fabrini2024,Wilken2023,GongAFM,Abraham_2024} and many simple or confined biological condensate structures~\cite{FERIC20161686,Chen2024,Wan2018,doyle2023,SANDERS2020306,Ye2025.05.14.654140,Wan2018,Fare2021,Zhao2024,Fei2017}, we show that the topological constraints change qualitatively between $N=2$, $N=3$ and $N\geq4$, in accordance to coloring rules. These constraints strongly affect the balance between Ostwald ripening and domain coalescence, thereby determining how the coarsening dynamics depend on the number of immiscible phases $N$. For instance, for the same material parameters, we can observe a 
hydrodynamic dominated regime for $N=2$, a crossover from hydrodynamic to diffusive dynamics for $N=3$, and diffusive-like scaling for $N\ge4$. By employing chromatic graph theory to estimate the frequency of coalescence events, we derive an analytical model that quantitatively captures the simulation results for $N\geq3$. In contrast, classical theories of Ostwald ripening based solely on diffusion underestimate the coarsening rate.

We further show that, by modifying the interfacial tensions, we can energetically penalize adjacencies between specific phases in the contact network. An interaction connectivity graph can be constructed to map the possible adjacencies, and depending on the resulting graph, different coarsening scaling laws can be encouraged for different phases in the same system due to a competition between diffusion and hydrodynamics. 
Finally, for non-planar phase organization, representative of unconfined systems, the space of possible configurations is significantly less constrained topologically. There is a change in coloring rules including no analogue of the four-color theorem. Thus, hydrodynamics is only suppressed gradually as the number of phases exceeds $N\gtrsim7$ instead of a sharp transition observed in the planar case. 

\section{Continuum Model for N-Component Phase Separation}~\label{sec:contmod}
To study the dynamics of many component fluid systems, we solve the following continuum equations of motion. First, we have the incompressible continuity and Navier-Stokes equations for conservation of mass and momentum,
\begin{eqnarray}
& \mathbf{\nabla}\cdot\mathbf{v} = 0,
\label{eqn:ce} \\
& \partial_t(\rho \mathbf{v}) + \mathbf{\nabla}\cdot(\rho\mathbf{vv})= - \mathbf{\nabla}\cdot\mathbf{P} + \eta\Delta \mathbf{v}.
\label{eqn:nse}
\end{eqnarray}
Here, $\rho$, $\mathbf{v}$, $\eta$ and $\mathbf{P}$ are the fluid density, velocity, dynamic viscosity, and pressure tensor, which includes the effects of interfacial tension. Second, the tracking of each fluid volume fraction $C_i$ is performed via $N-1$ Cahn-Hilliard equations,
\begin{equation}
\label{eq:CHFlux}
\partial_t C_i+\nabla\cdot(\mathbf{v}C_i)= M\mathbf{\nabla}^2\sum^N_{j=1}\alpha_{ij}(\mu_j+\phi_j+\xi_j),
\end{equation}
with $M$ a constant mobility parameter. The evolution of the $N^{th}$ immiscible component is implicit from the constraint that $C_N=1-\sum_{i=1}^{N-1}C_i$. Technical details of the lattice Boltzmann implementation of Eqs. \ref{eqn:ce}-\ref{eq:CHFlux} are given in Appendix~\ref{sec:LBE}.

The chemical potential $\mu_j$ and pressure tensor $\bf{P}$ can be derived from the fluid mixing free energy $F$. For a system of $N$ immiscible components, the following free energy functional allows independent selection of the interfacial tension between each immiscible component~\cite{boyerncomp,dongncomp},
\begin{align}
F &= \int \left[ E_b + E_i \right] dV, \\
E_b &= \sum_{i\neq j}^N \frac{\beta_{ij}}{2} \Big[ f(C_i) + f(C_j) - f(C_i + C_j) \Big], \notag \\
f(C_i) &= C_i^2 (1-C_i)^2, \quad E_i = -\sum_{i\neq j}^N \frac{\kappa_{ij}}{2} \nabla C_i \cdot \nabla C_j. \notag
\label{eqn:FreeEnergy}
\end{align}
$E_b$ is a bulk contribution and ensures that, far from the interface, the energy is minimised when $C_i$ is either $1$ (present) or $0$ (absent) for each immiscible component. The contribution $E_i$ leads to a diffuse interface with width $D=\sqrt{8\kappa_{ij}/\beta_{ij}}$ between fluid regions and ensures that the excess free energy per unit area of the interface corresponds to the desired interfacial tensions $\sigma_{ij}=\sqrt{2\kappa_{ij}\beta_{ij}/9}$. In this work, we set $D=3$ (in lattice units) and vary $\kappa_{ij}\beta_{ij}$ to tune the fluid-fluid interfacial tensions. The chemical potential is then calculated as $\mu_j=\partial F/\partial C_j$ and the pressure tensor is given by
\begin{equation}
\label{eq:stress_tensor_compact}
\mathbf{P}
= \left[p_h - E_b - E_i\right]\mathbf{I}
- \sum_{i\neq j}\kappa_{ij}\,\nabla C_i \otimes \nabla C_j,
\end{equation}
where $p_h$ is the hydrodynamic pressure.

An important feature of our lattice Boltzmann scheme is its reduction consistency property: an initially absent component cannot erroneously nucleate. This issue becomes increasingly prominent with the number of fluid components, and can lead to vastly inaccurate results or numerical instabilities \cite{NCompMethod,boyerncomp,dongncomp}, as discussed in Appendix~\ref{sec:reduct}. The $\phi_j$ term in Eq.~\ref{eq:CHFlux} is key for reduction consistency, and it takes the following form
\begin{equation}
\phi_j=\frac{12}{D}\sum_{\substack{1\leq k<l<m\leq N\\k\neq j, l\neq j, m\neq j}}\Lambda_{j;{j,k,l,m}}C_kC_lC_m. \label{eq:G} 
\end{equation}
Finally, an optional stabilizing term $\xi_j$ is included to ensure the free energy remains bounded from below outside the physical range $C_j\in[0,1]$,
\begin{equation}
\xi_j=\frac{12\Omega}{D}\sum_{\substack{1\leq k<l\leq N\\k\neq j, l\neq j}}C_k^2C_l^2\left(2C_j-\sum_{\substack{1\leq m\leq N\\m\neq j, k, l}}\Theta_{kl;j;m}C_m\right), \label{eq:P}
\end{equation}
where $\Omega$ is a free parameter to control the strength of the stabilizing contribution. This term is useful for cases with positive spreading coefficients, leading to cloaking of fluid domains, as discussed in Sec.~\ref{sec:energy}.

A description of the procedures to evaluate coefficients $\alpha_{ij}$, $\Lambda_{j;{j,k,l,m}}$ and $\Theta_{kl;j;m}$~\cite{boyerncomp} are provided in Appendix~\ref{sec:reduct}.
Unless specified otherwise, we set $\sigma_{ij}=0.005$, $\rho=1$, $\eta=1/6$, $\Omega=0$ and $M=1/300$ in simulation units.

\begin{figure*}[!t]
	\centering
	\includegraphics[width=\textwidth]{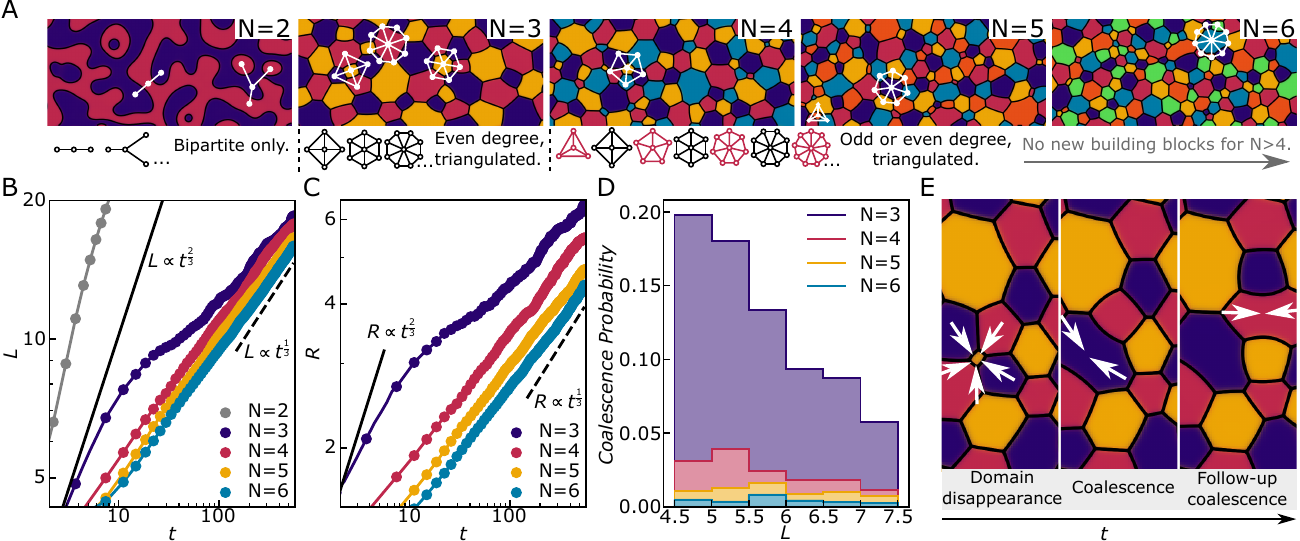}
	\caption{
    Phase separation dynamics in the planar case for $N=2$ to $N=6$.
    (A) Subsets of the fluid structure for $N=2$ at $t=1.85$ and $N=3$, $4$, $5$ and $6$ at $t=139$. Insets: Topologically distinct building blocks that make up the corresponding contact networks, determined using the approach in Appendix~\ref{sec:contactnetwork}. (B) $L$ over time for varying numbers of components in a periodic domain with dimensions $94.3\sqrt{N}\times94.3\sqrt{N}$. (C) Domain radius $R$ over time for $N\geq3$, measured from the same simulations in panel B. (D) Probability that an individual fluid region coalesces with a neighbour in a time interval of $\Delta t=1$ for a given $L$ with varying numbers of components. (E) For $N>2$, coarsening proceeds through diffusive Ostwald ripening, followed by coalescence when a domain disappears. These coalescence events can lead to follow-up coalescence when the surrounding domains move.}
	\label{fig:2D}
\end{figure*}

\section{Results}~\label{sec:results}

\subsection{Universal dynamic scaling behavior for $N>3$ phase separation}~\label{sec:planar} 
We start by considering the influence of the number of fluid phase $N$ on the coarsening dynamics and permissible phase structures in the planar case, where the contact network can be projected onto a plane without edge crossings. This is representative for a number of simple biological structures~\cite{FERIC20161686,Chen2024,Wan2018,doyle2023,SANDERS2020306,Ye2025.05.14.654140,Wan2018,Fare2021,Zhao2024,Fei2017} and synthetic droplet monolayers~\cite{chaderjian2025diversedistinctdenselypacked,Fabrini2024,Wilken2023,GongAFM,Abraham_2024} since they are often confined or assembled in thin monolayers. To understand phase separation in the planar case, it is sufficient to conduct the simulations in two dimensions, and for simplicity, we will consider equal interfacial tensions and volume fractions for all fluid components. The generalizations to different interfacial tensions and volume fractions are provided in Sec.~\ref{sec:energy}, while three-dimensional simulations are shown in Sec.~\ref{sec:nonplanar}.

Snapshots of the simulation results for $N = 2$ to $N=6$ are shown in Fig.~\ref{fig:2D}(A). To initialize the simulation, we evaluate a random decimal between $0$ and $2/N$ from a uniform distribution for each fluid on every lattice point, and normalize these so that the sum is equal to $\sum C_i=1$. We apply periodic boundaries and use a system size that scales with $N$ to ensure that the volume of each fluid in the periodic unit is fixed between simulations.

Given the fluid phase morphologies, we can construct contact networks, as shown in the insets of Fig.~\ref{fig:2D}(A). They are identified using the approach in Appendix~\ref{sec:contactnetwork}, and the topologically distinct building blocks correspond to subgraphs that are colorable with $N$ colors. Simple examples of three and four colorable contact networks were given in Fig.~\ref{fig:fourcol}(B). For $N\geq3$, the contact networks are naturally triangulated, because the system preferentially forms triple points that balance the interfacial tensions. For $N=3$ only even degree contact networks are permissible, while even and odd degree contact networks are possble for $N\ge 4$. Moreover, we observe no change in the building blocks for $N>4$, consistent with the four-color theorem.

To quantify the coarsening dynamics, we compute the characteristic length scale for a given configuration in two different ways. First, to obtain a characteristic length scale irrespective of the fluid morphology, we can use the structure factor to calculate~\cite{phaseseptheo2,phaseseptheo3}
\begin{equation}
L_i=2\pi\frac{\int S_i(k) d^nk}{\int kS_i(k) d^nk},\quad\forall i\in \{1,...,N\},
\label{eq:L}
\end{equation}
where $n$ is the spatial dimension of the system. The spherically averaged structure factor $S_i(k)=\left<\hat\lambda_i(\mathbf{k})\hat\lambda_i(-\mathbf{k})\right>$ is given in terms of Fourier transforms of $\lambda_i(\mathbf{r})=H(C_i(\mathbf{r})-0.5)$, where $H$ is the Heaviside step function and $\mathbf{r}$ is the position vector. The characteristic length $L$ is the average of $L_i$ over all components, and is a weighted average over all length scales present in the system. Second, when domains are isolated and roughly isotropic, we can approximate the average domain radius as~\cite{WagnerYeomans} 
\begin{equation}
    R=\sqrt{A/\pi n_{\mathrm{d}}},
\end{equation}
where $A$ is the area of the periodic unit and $n_{\mathrm{d}}$ is the total number of fluid domains in the unit. In contrast to $L$ from the structure factor, $R$ only contains information about the average domain radius and is not influenced by the average spacing between domains of the same phase. We demonstrate this in Fig. S1, where we measure $L$ and $R$ for a fixed domain radius and varying domain spacing. 
We also non-dimensionalize the length and time by $L^*=L/D$, $R^*=R/D$ and $t^*=tM/D^2$, where $M$ is the mobility parameter and $D$ is the diffuse interface width. From here on, we always present the non-dimensionalized quantities and drop the superscript.

The resulting coarsening dynamics can be understood by considering the space of colorable contact networks for a given $N$, together with the interfacial tensions that select hydrodynamically stable configurations. For $N=2$, previous works have shown that the coarsening dynamics can be in the diffusive ($L\propto t^{1/3}$ ~\cite{ostwald,marqusee}), viscous hydrodynamics ($L\propto t$), or inertial hydrodynamics ($L\propto t^{2/3}$) regime~\cite{phasesepscaling,phasesepscalingsource,Wagner_2001,WagnerYeomans}. Here, we choose the simulation parameters such that the bicontinuous structures coarsen hydrodynamically, driven by interfacial tension to minimize their interfacial area, and in turn investigate how increasing $N$ changes the phase separation dynamics. We clearly observe a scaling law of $L\propto t^{2/3}$ for $N=2$, as shown in Fig.~\ref{fig:2D}(B).

Increasing to $N\ge 3$ leads to the formation of triple points which temporarily arrest the hydrodynamic driving force. Coarsening then proceeds via a combination of Ostwald ripening and domain coalescence. Ostwald ripening is a diffusive process in which larger domains grow and smaller ones shrink and eventually disappear. This domain disappearance event can lead to a cascade of coalescence events, where adjacent domains of the same colors are brought into contact (see Movie S1).

The total timescale $t_{\mathrm{char}}$ for coarsening depends on the time taken for a domain to disappear through Ostwald ripening $t_{\mathrm{d}}$, and the time taken for subsequent coalescence events $t_{\mathrm{c}}$,
\begin{equation}
t_{\mathrm{char}} = t_{\mathrm{d}} + t_{\mathrm{c}}.
\end{equation}
Considering the results for $N=3$, we find that at early times (small characteristic lengths), $t_{\mathrm{c}}$ is the bottleneck, and coarsening is dominated by frequent coalescence events between multiple regions (Movie S2). Following the results for $N=2$, we thus expect an initial scaling of $L\propto t^{2/3}$ and $R\propto t^{2/3}$, as shown in Fig.~\ref{fig:2D}(B,C). At late times (large characteristic lengths), $t_{\mathrm{d}}$ is the slow timescale, leading to a scaling of $L\propto t^{1/3}$ characteristic of Ostwald ripening~\cite{ostwald,marqusee}. We also obtain a similar late time scaling for the domain radius of $R\propto t^{1/3}$ in Fig.~\ref{fig:2D}(C).

Such crossover is not present for $N \ge 4$. As shown in Fig.~\ref{fig:2D}(B,C), we observe an approach to $L\propto t^{1/3}$ and $R\propto t^{1/3}$ scalings throughout the simulations. 
Coalescence events are significantly less probable for $N \ge 4$ than for $N=3$ at small $L$, see Fig.~\ref{fig:2D}(D), which we measure using the protocol in Appendix~\ref{sec:coalesce}. This means $t_{\mathrm{c}}$ is never the bottleneck. To explain this, first, the average volume fraction per unit area scales with $1/N$; therefore, the probability that multiple regions of the same fluid are close enough to undergo coalescence decreases with $N$. Second, coalescence events are not independent. As illustrated in Fig.~\ref{fig:2D}(E) and Movie S{\color{blue}1}, it is possible that the coalescence of two domains (purple domains) gives rise to further coalescence (red domains). This effect is strongest for $N=3$, because every Ostwald ripening event must lead to at least one coalescence event to maintain the even neighbor structure. In contrast, with $N\geq4$, the four-color theorem allows both even and odd neighbor structures and each Ostwald ripening event is not guaranteed to cause coalescence.
Nonetheless, as we will discuss in the next section, the coarsening dynamics cannot be predicted accurately if we only account for Ostwald ripening and ignore coalescence altogether. Coalescence events accelerate domain growth.

\subsection{Theoretical model for the coarsening dynamics}\label{sec:model}

To obtain a more quantitative description of the scaling behavior for $N>2$, we construct a minimal model for how the rate of coarsening at late times depends on $N$. 
For binary Ostwald ripening~\cite{marqusee}, the change in the domain radius $R$ can be written as
\begin{equation}
\label{eq:drdta}
\frac{d{R}}{dt} \propto \frac{a_N^3}{R^2},
\end{equation}
where we generalize the coefficient $a_N$ to depend on the number of components. We can calculate $a_N$ for different $N$ using the approach in Appendix~\ref{sec:aN}, leading to $a_3=0.975$, $a_4=0.923$, $a_5=0.887$ and $a_6=0.859$.

During the diffusive process, domain disappearance and coalescence will occur, which will accelerate coarsening. Because coalescence and domain disappearance are discrete events, we consider on average how the number of regions is changing with time. For concreteness, we use our definition for the average domain size, as defined in the previous sub-section, such that the number of fluid domains is $n_{\mathrm{d}} = A/\pi R^2$. Correspondingly,
\begin{equation}
\label{eq:dndt}
\frac{dn_{\mathrm{d}}}{dt}=\frac{dn_{\mathrm{d}}}{dR}\frac{dR}{dt} \propto -\frac{2a_N^3\pi^{3/2}}{A^{3/2}} n^{5/2}.
\end{equation}
This can be interpreted as the expected rate of domain disappearance due to Ostwald ripening. Each Ostwald ripening disappearance event reduces the number of domains by one. If an Ostwald ripening disappearance leads to, on average, $\langle n_c\rangle$ coalescence events (correspondingly, $\langle n_c\rangle$ reduction in the number of domains), then the total rate of decrease for the number of domains is enhanced by a factor of $1+\langle n_c\rangle$,
\begin{equation}
\label{eq:dndt}
\frac{dn}{dt} \propto -\frac{2a_N^3\pi^{3/2}}{A^{3/2}}(1+\left<n_c\right>)n^{5/2}.
\end{equation}
Correspondingly, $R$ changes with time as
\begin{eqnarray}
\frac{dR}{dt} \propto \frac{a_N^3(1+\left<n_c\right>)}{R^2}, \nonumber \\
R \propto a_N(1+\left<n_c\right>)^{\frac{1}{3}}t^{\frac{1}{3}}. \label{eq:scaling}
\end{eqnarray}

\begin{figure*}[!t]
	\centering
	\includegraphics[width=\textwidth]{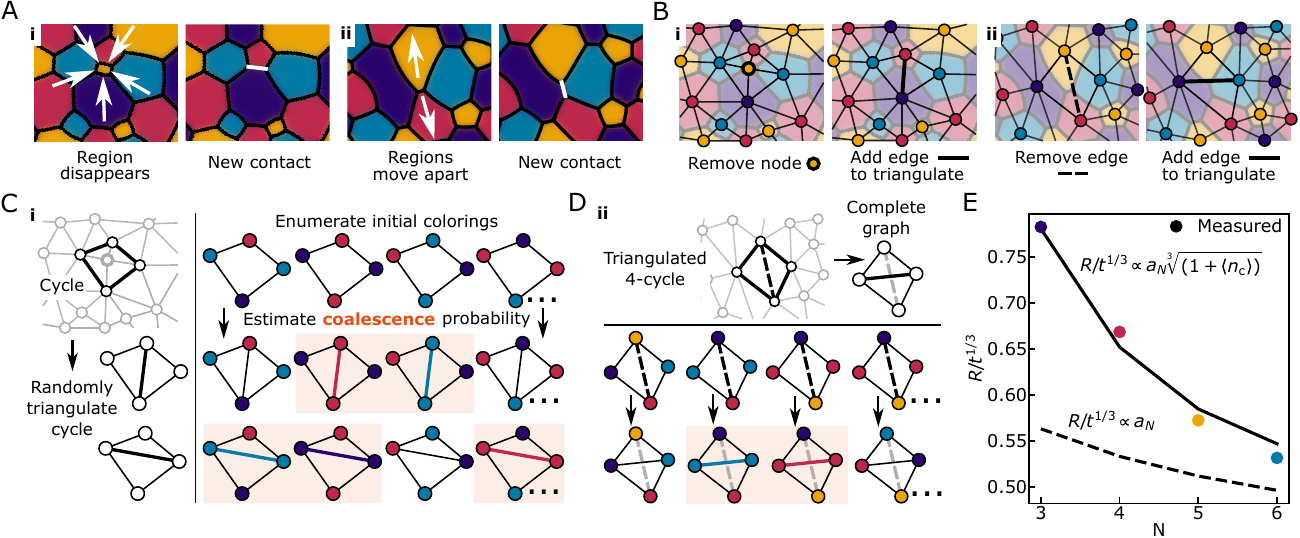}
	\caption{
    Employing graph coloring theorems to develop scaling theory for the coarsening dynamics. (A) New contacts between fluid domains can be created through two mechanisms (i) and (ii) shown. (B) The mechanisms in panel A can be understood as the removal of (i) a node or (ii) an edge from the contact network, followed by the addition of new edges to satisfy the constraint that the network is triangulated. (C) Following the removal of a node in (i), the surrounding nodes form a cycle, and the addition of edges triangulates the cycle. The probability of coalescence can be estimated for each edge by enumerating all unique initial cycle colorings. (D) Following edge removal in (ii), both the initial and new edge must connect nodes of different colors to avoid coalescence, forming a complete graph of four nodes. (E) Dependence of the prefactor in the scaling of $R\propto t^{1/3}$ in Fig. 2(C) on $N$. The expected scaling for our theory and the classical Ostwald ripening~\cite{marqusee} are shown as comparison.}
	\label{fig:graph}
\end{figure*}

To evaluate $\left<n_c\right>$, we need to estimate how new contacts between adjacent domains are generated. As shown in Fig.~\ref{fig:graph}(A), there are two key mechanisms: (i) After a fluid domain disappears, some surrounding fluid regions must meet and form new contacts; (ii) Upon coalescence, the resulting rearrangement can lead to new contacts. We can equivalently express these mechanisms as changes in the topology of the contact network.
In the contact network, we can understand (i) as the removal of a node when a region disappears and (ii) as the removal of an edge when two regions move apart. Following each of these, the new edges that are created must respect the constraint that the contact network is triangulated to balance the interfacial tensions, illustrated in Fig.~\ref{fig:graph}(B). 

We can split $\left<n_{\mathrm{c}}\right>$ into the expected number of coalescence events following domain disappearance from (i), $\left<n_{\mathrm{c|(i)}}\right>$, and from (ii), $\left<n_{\mathrm{c|(ii)}}\right>$, and consider each case separately
\begin{equation}
\label{eq:ncsplit}
\left<n_{\mathrm{c}}\right>=\left<n_{\mathrm{c|(i)}}\right>+\left<n_{\mathrm{c|(ii)}}\right>\left<n_{\mathrm{(ii)}}\right>,
\end{equation}
where $\left<n_{\mathrm{(ii)}}\right>$ is the expected number of edge removal events per domain disappearance event.

First, we estimate $\left<n_{\mathrm{c|(i)}}\right>$. Consider a node that is about to be removed from the contact network. This central node has edges connecting it to its immediate neighboring nodes, and those neighboring nodes connect to each other in a cycle, shown in Fig.~\ref{fig:graph}(C). We assume the cycle graph is randomly colored with $N-1$ colors, because the $N^{\textrm{th}}$ color is occupied by the removed node, with no edges connecting nodes of the same color. When the cental node and its associated edges are removed, the other nodes in the cycle must now form new edges between themselves, while ensuring that the resulting graph is triangulated. Importantly, if the triangulation connects two nodes of the same color, coalescence will occur. We assume that the initial coloring is randomly chosen from the set of valid colorings of a cycle, and the triangulation is randomly chosen from the set of valid triangulations for the cycle, as illustrated in Fig.~\ref{fig:graph}(C).

Each added edge in the triangulation can lead to at most one coalescence event. The expected number of total coalescence events is given by the following sum over possible added edges $e$
\begin{equation}
\left<n_{\mathrm{c|(i)}}\right>=\sum_e(\mathcal{P}_e\times \mathcal{P}_{\mathrm{c}|e}),
\end{equation}
where $\mathcal{P}_e$ is the probability that a given edge appears in the random triangulation and $\mathcal{P}_{\mathrm{c}|e}$ is the probability that the edge connects two nodes of the same color (leading to coalescence). For a given edge added to a cycle with $m$ nodes, $\mathcal{P}_{\mathrm{c}|e}$ can be evaluated by counting the number of initial colorings of the cycle where two nodes of the same color would be connected once the edge is added, divided by the total number of possible initial colorings, 
\begin{equation}
\mathcal{P}_{\mathrm{c}|e}=\frac{p_{\mathrm{cyl}}(N-1,m)-p_{\mathrm{cyl}+e}(N-1,m)}{p_{\mathrm{ cyl}}(N-1,m)}.
\end{equation}
Here, $p_{\mathrm{cyl}}(N-1,m)$ denotes the number of unique ways to color a cycle with $N-1$ colors and $m$ nodes, and $p_{\mathrm{cyl}+e}(N-1,m)$ provides the number of ways to color the cycle with the additional constraint that nodes connected by the new edge $e$ must have different colors. The difference between these numbers is the number of cycle colorings where nodes connected by the added edge would share the same color, leading to coalescence. In graph theory, the numbers $p_{\mathrm{cyl}}(\mathcal{C},m)$ and $p_{\mathrm{cyl}+e}(\mathcal{C},m)$ can be evaluated from chromatic polynomials,
where $\mathcal{C}$ is the number of colors~\cite{dong2005chromatic}. 

As disappearance only occurs for small domains, we measure the average number of neighbors for the smallest $5\%$ of domain areas across all $N \ge 3$ simulations, which correspond to $\approx4.14$ (late time average between $t=370$ and $t=556$). Taking the $m=4$ cycle, there are two possible added edges with $\mathcal{P}_e=0.5$, each of which fully triangulates the cycle. The relevant chromatic polynomials for $m=4$ are~\cite{dong2005chromatic}
\begin{eqnarray}
&p_{\mathrm{cyl}}(\mathcal{C},m)=(\mathcal{C}-1)^{m}+(-1)^{m}(\mathcal{C}-1), \\
&p_{\mathrm{cyl}+e}(\mathcal{C},m)=p_{\mathrm{tri}}(\mathcal{C},m)=\mathcal{C}(\mathcal{C}-1)(\mathcal{C}-2)^{m-2},
\end{eqnarray}
where $p_{\mathrm{tri}}$ provides the number of ways to color a triangulated polygon without connected nodes sharing a color.
The expected number of coalescence events is thus
\begin{eqnarray}
\label{eq:nci}
\left<n_{\mathrm{c|(i)}}\right>&=&2\times0.5\times\left(\frac{p_{\mathrm{cyl}}(N-1,4)-p_{\mathrm{tri}}(N-1,4)}{p_{\mathrm{cyl}}(N-1,4)}\right) \nonumber \\
&=&\frac{N-2}{N^2-5N+7}.
\end{eqnarray}

Now, we consider $\left<n_{\mathrm{c|(ii)}}\right>$ and $\left<n_{\mathrm{(ii)}}\right>$. 
We assume that edge removal primarily occurs when domains move due to coalescence, and that the number of edge removal events is proportional to the number of coalescence events such that 
\begin{equation}
\left<n_{\mathrm{(ii)}}\right>=k_{\mathrm{e}}\left<n_{\mathrm{c}}\right>. \label{eq:ke}
\end{equation} 
We estimate $k_{\mathrm{e}}\approx0.402$ empirically from our data by measuring the rate of coalescence at late times, following the approach in Appendix~\ref{sec:ke}. To evaluate $\left<n_{\mathrm{c|(ii)}}\right>$, we note that, if an edge is removed, the contact network is always retriangulated by inserting a new chord (an edge joining two non-adjacent vertices in the cycle) within the cycle of 4 nodes for which the removed edge was a chord, shown in Fig.~\ref{fig:graph}(D). The only initial colorings of the triangulated 4-cycle that do not lead to coalescence are those that use four different colors, because both the initial and new chord must connect different colors.
The number of colorings of the 4-cycle that would not lead to coalescence is given by the chromatic polynomial for a complete graph~\cite{dong2005chromatic}
\begin{equation}
p_{\mathrm{complete}}(\mathcal{C},m)=\mathcal{C}(\mathcal{C}-1)(\mathcal{C}-2)...(\mathcal{C}-m+1).
\end{equation}
Similar to before, the expected number of coalescence events $\left<n_{\mathrm{c|(ii)}}\right>$ can be evaluated by counting the number of initial colorings of the triangulated 4-cycle where two nodes of the same color would be connected once the other chord is added, divided by the total number of possible initial colorings 
\begin{eqnarray}
\label{eq:nce}
\left<n_{\mathrm{c|(ii)}}\right>&=&\left(\frac{p_{\mathrm{tri}}(N,4)-p_{\mathrm{complete}}(N,4)}{p_{\mathrm{tri}}(N,4)}\right)\nonumber \\
&=&\frac{1}{N-2}.
\end{eqnarray}

Bringing the results together, we predict 
\begin{equation}
\label{eq:ncprop}
\left<n_{\mathrm{c}}\right>=\frac{N-2}{(N^2-5N+7)(1-k_{\mathrm{e}}/(N-2))}.
\end{equation}
Importantly, we can verify this prediction by comparing the average of $R/t^{1/3}$ at late times (between $t=370$ and $t=556$) from Fig.~\ref{fig:2D}(D) against $a_N(1+\left<n_c\right>)^{\frac{1}{3}}$ for varying $N$. The results are shown in Fig.~\ref{fig:2D}(E) where we have fitted for the constant of proportionality in Eq.~\ref{eq:scaling}. We observe that the dependency on $N$ is well captured. For comparison, we also show the expected result from classical Ostwald ripening model~\cite{marqusee}, $a_N$, which does not include contributions due to domain coalescence ($1+\langle n_c \rangle$). It is clear that the $N$-dependence is poorly represented. Due to coalescence, the resulting coarsening dynamics is faster compared to that driven only by diffusion, and our model accurately captures the numerical results.

\begin{figure*}[!t]
	\centering
	\includegraphics[width=\textwidth]{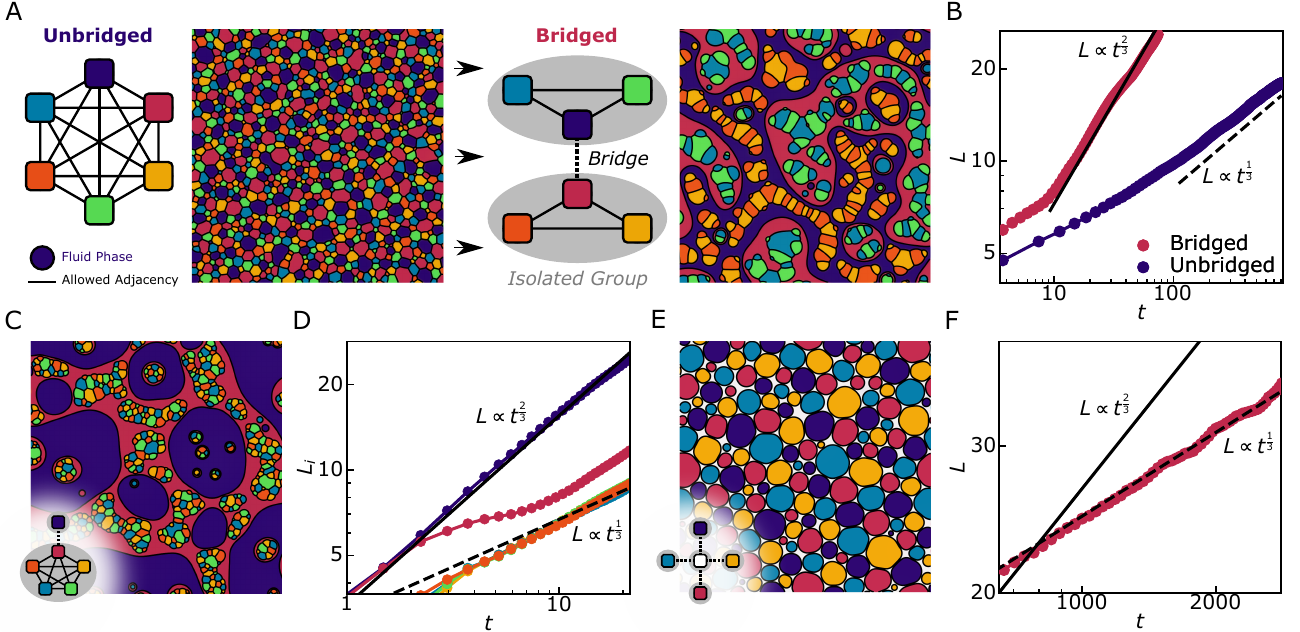}
    \caption{The dominant coarsening dynamics can be modulated through bridged interaction connectivity graphs. (A) Fluid structure during $N=6$ phase separation in a $200\times200$ domain for a bridged and an unbridged interfacial tension configuration (Movie S3, 4). Interfacial tensions and volume fractions are given in Table S1. (B) $L$ over time for the bridged and unbridged interfacial tension configurations in panel A. (C) Fluid structure and interaction connectivity graph for $N=6$ phase separation in a $267\times267$ domain where the red fluid cloaks four of the others (Movie S5). The purple fluid only contacts the red fluid. Interfacial tensions and volume fractions are given in the Table S2. (D) $L_i$ over time with the configuration in panel C, now plotted individually as the average length scale of each phase $i$ in the system. (E) Fluid structure and interaction connectivity graph for $N=5$ phase separation in a $333\times333$ domain where the white fluid cloaks the interface between all other fluid phases (Movie S6). Interfacial tensions and volume fractions are given in the Table S3. (F) $L$ over time for the configuration in panel E. Fluctuations occur due to finite size effects. We focus on late times from the point at which all domains become fully isolated, which is delayed due to resolution constraints of thin white regions and the finite interface width. 
    }
	\label{fig:PhaseSep}
\end{figure*}

\subsection{Energetic penalties to phase adjacency control the dominant coarsening dynamics}~\label{sec:energy} We next consider whether the suppression of hydrodynamic scaling is robust to physical constraints on adjacencies in the contact network. To investigate this, we vary the fluid interfacial tensions with both positive and negative spreading parameters $S_{ij;k}=\sigma_{ij}-\sigma_{ik}-\sigma_{jk}$. If $S_{ij;k}$ is positive, it is always energetically preferable for fluid $k$ to spread at the interface between $i$ and $j$ to prevent adjacency between them, and triple points that include $i$, $j$ and $k$ can no longer form. We represent the energetically permissible adjacencies with interaction connectivity graphs consisting of $N$ colored nodes, as defined by Mao et al.~\cite{phaseseptheo1}. A connection between two nodes indicates that adjacency is allowed between fluids of those colors. To draw these, we start with a fully connected graph. For an arbitrary triplet of nodes $ijk$, we iterate over the spreading parameters $S_{ij;k}$. For any positive spreading parameter, we delete the connection between nodes $i$ and $j$. We then repeat this with all possible combinations of node triplets to complete the graph. Once the connections are mapped on the graph, we identify bridges (drawn as a dashed line in this work) as any connection that is not involved in any cycles on the graph and whose deletion would produce two isolated subgraphs. This procedure is described visually in Fig. S2.

Fig.~\ref{fig:PhaseSep}(A) compares an unbridged and a bridged interaction connectivity graph for $N=6$, with the specific interfacial tensions and volume fractions given in Table S1. As before, the unbridged case is dominated by $L \propto t^{1/3}$ coarsening dynamics (Fig.~\ref{fig:PhaseSep}(B)), which is consistent across a wide range of different interfacial tension and volume fraction configurations (Supplemental Material, Section IV). In contrast, for the bridged case in Fig.~\ref{fig:PhaseSep}(A), since there is a single bridge between two sub-groups of nodes, triple points cannot form between the two isolated sub-groups of fluids and the contact network is no longer fully triangulated. This lack of stabilising triple points leads to a bicontinuous interface, and the phase structure is effectively bipartite between the two isolated groups of fluids. As a result, the fluids readily redistribute throughout the domain and hydrodynamic coarsening is restored with $L\propto t^{2/3}$ (Fig.~\ref{fig:PhaseSep}(B)), similar to the $N=2$ case. 

Moreover, by tuning the energetically allowed triple points we can decouple the coarsening dynamics within each side of a bridge and intentionally disrupt the self similarity of the global structure, promoting different dominant coarsening mechanisms for the individual fluid components. For example, this is shown in Fig.~\ref{fig:PhaseSep}(C), where we configure the interaction connectivity graph to induce four fluid phases (orange, yellow, light blue and green) encapsulated within one phase (red) of an otherwise binary mixture (red and purple). The encapsulated fluids form stable clusters and primarily evolve through diffusive Ostwald ripening, while the purple fluid coarsens hydrodynamically (Fig.~\ref{fig:PhaseSep}(D)). As the red phase simultaneously interacts with the surrounding purple fluid and with the smaller clusters contained within it, the coarsening is influenced by both diffusion and hydrodynamics.

In the absence of stabilising triple points, we can reintroduce arrested hydrodynamics through mechanical jamming. We demonstrate this through an interaction connectivity graph that contains four or more bridges that extend from a single intermediary fluid, leading to four isolated fluid domains separated by the intermediary fluid, as shown in Fig.~\ref{fig:PhaseSep}(E). Despite the forbidden adjacencies between fluid phases, isolated domains are packed very tightly and can become pseudo-adjacent through a thin layer of the intermediary fluid. The packing morphology qualitatively resembles the unbridged case and the four-color theorem again applies to permit structures that avoid pseudo-adjacency of like fluids, leading to arrested hydrodynamics due to jamming of the domains (Fig.~\ref{fig:PhaseSep}(F)). This allows for stable dense packings of isolated droplets, which are consistent with recent observations in experiments with DNA nanostar condensates~\cite{chaderjian2025diversedistinctdenselypacked}.

\subsection{Geometric confinement modulates planar and non-planar coarsening dynamics}~\label{sec:nonplanar} 
We now consider phase separation dynamics in three dimensions, where the contact networks are in general non-planar. This changes the coloring rules, since there is no equivalent to the four-color theorem guaranteeing that all contact networks are colorable. For simplicity, we again use equal interfacial tensions and volume fractions for simplicity, and study the phase separation dynamics with up to $N=8$ components. 

\begin{figure}[!t]
	\centering
	\includegraphics[width=\columnwidth]{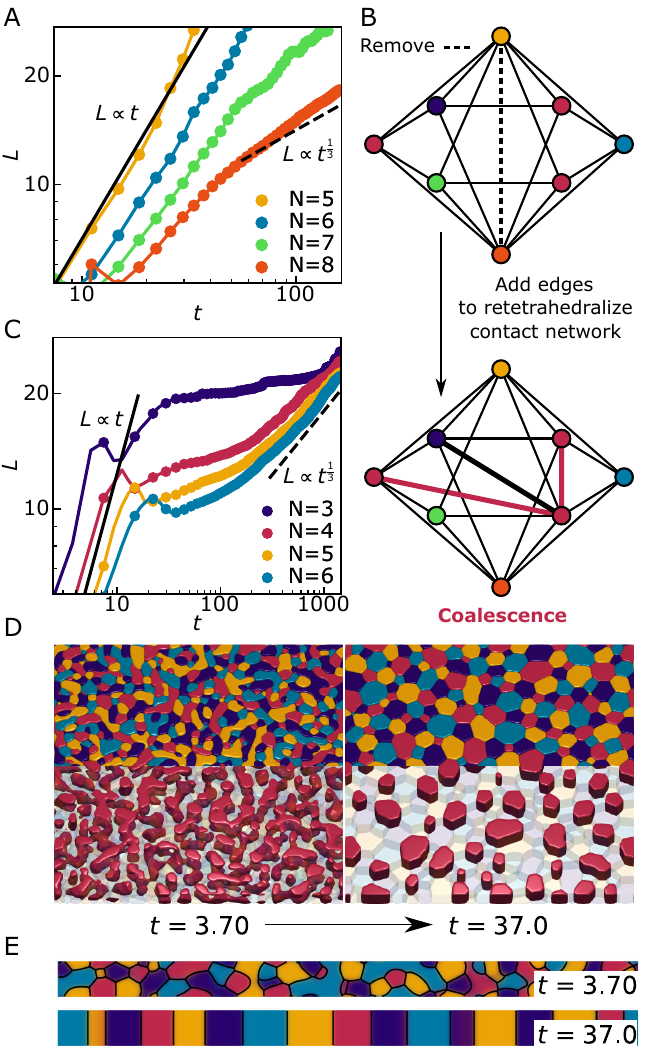}
	\caption{For non-planar contact networks, hydrodynamic scaling persists for larger $N$, but planar hydrodynamic suppression can be recovered through confinement. (A) $L$ over time in a cubic domain of $27.1\sqrt[3]{N}\times27.1\sqrt[3]{N}\times27.1\sqrt[3]{N}$ for $N=5$, $6$, $7$ and $8$. (B) Illustration showing that triangulating a planar contact network after edge removal leads to at most one coalescence event, while retetrahedralizing a non-planar contact network after edge removal can lead to multiple coalescence events. (C) $L$ over time in a thin domain of $67.5\sqrt{N}\times67.5\sqrt{N}\times 8.33$ for $N=3$, $4$, $5$ and $6$. (D) Three-dimensional visualisation of the thin phase structure for $N=4$ at $t=3.70$ and $t=37.0$ (see Movies S7 and S8 for full time evolution), with three fluids made transparent in the lower half of the images to highlight the change in topology over time. (E) Side view of the thin film phase structure for $N=4$ at $t=3.70$ and $t=37.0$ (see Movie S9 for full time evolution).}
	\label{fig:3D}
\end{figure}

A key consequence for the non-planar contact network is that we no longer observe a strict upper bound in the number of concurrent phases needed to arrest hydrodynamic coarsening. As shown in Fig.~\ref{fig:3D}(A), systems with $N=5$ and $N=6$ follow viscous hydrodynamics with $L \propto t$~\cite{phasesepscalingsource}, with a gradual transition toward diffusive $L \propto t^{1/3}$ scaling as the number of components is increased further. This transition occurs because the probability of like phases coming into contact for coalescence decreases asymptotically for large $N$, which increases the role of diffusion in the coarsening process.

Domain coalescence is significantly more dominant in the non-planar case for several reasons. Firstly, the average number of neighbors per node is much greater than the planar case in Section~\ref{sec:planar}. For example, this is $\approx14.6$ for $N=6$ in three dimensions compared to $\approx 6$ for $N=6$ in two dimensions. This means a significant number of new edges need to be added following domain disappearance, leading to more opportunities for coalescence. Secondly, edge removal is no longer guaranteed to add just one new edge. The non-planar contact network is tetrahedralized, and retetrahedralizing the network following edge removal can lead to multiple opportunities for coalescence, as shown in Fig.~\ref{fig:3D}(B). Consequently, phase structures that temporarily arrest phase separation are harder to realize and the influence of hydrodynamics is preserved for much larger $N$.

We can transition between non-planar and planar coarsening dynamics by confining the phase separation process to a thin domain, as illustrated in Fig.~\ref{fig:3D}(C-E)). Confinement is pervasive in biological cells, either externally by surrounding tissue geometry and mechanical compression, or internally through fibrillar networks and membranes~\cite{quiroz,ZhaoConfine2024,Liu2023,Snead2022}. Furthermore, dense packings of synthetic DNA droplets are often assembled in thin monolayers~\cite{chaderjian2025diversedistinctdenselypacked,Wilken2023}. Here, we calculate $L$ in Eq.~\ref{eq:L} using the circularly averaged structure factor of a single $67.5\sqrt{N}\times67.5\sqrt{N}$ slice through the domain, to allow direct comparison with the two-dimensional scenario in Section~\ref{sec:planar}. For the thin domain case, coarsening initially follows viscous hydrodynamics until the average fluid domain size is larger than the film thickness, following which diffusion becomes the dominant effect and the curves approach $L \propto t^{1/3}$  (Fig.~\ref{fig:3D}(C), results for other film thicknesses are plotted together in Fig. S3). This happens due to the initially complex three-dimensional phase morphology eventually becoming uniform in the confining dimension when the domains outgrow the film thickness (Fig.~\ref{fig:3D}(D,E)), restricting the system to planar contact networks and reintroducing the coarsening behavior in Section~\ref{sec:planar}. The dip in $L$ around $t=10$ in Fig.~\ref{fig:3D}(C) occurs due to the change in morphology from elongated continuous domains to the more isotropic structure observed in Fig.~\ref{fig:3D}(D), as the measure of $L$ using Eq.~\ref{eq:L} responds to changes in the characteristic morphology.

\section{Discussion}~\label{sec:discussion}

The phase separation dynamics of many component fluid mixtures is central to spatial organization in many living and synthetic systems. Here, we develop a contact network framework that establishes a natural connection from mathematical coloring problems to the phase separation dynamics in these systems. Our findings highlight three key topological features. First, the space of contact networks that can be colored by the number of phases determines the probability of hydrodynamic coalescence following a topological change, such as domain disappearance due to Ostwald ripening or a previous domain coalescence. Second, the permissible phase adjacencies imposed by the interfacial tensions controls whether phases can form bicontinuous interfaces that coarsen hydrodynamically to minimise their interfacial area. Third, the ability for the system to adopt planar or non-planar contact networks is determined by the system geometry, with non-planar contact networks creating additional opportunities for hydrodynamic coalescence following a topological change. 

Taken together, these topological features lead to several important consequences. For planar phase organization without interface cloaking, we are able to derive a universal diffusive-like scaling law for $N\ge 3$. While domain coalescence events are infrequent, they are important to accelerate domain growth. When there is interface cloaking, corresponding to bridges in the interaction connectivity graphs, we show that hydrodynamic scaling can be reintroduced, and unique coarsening pathways can be realized for different fluids in the same system. In addition, we show that confinement - or its lack thereof - changes the nature of the contact networks and hence the coarsening behaviors. For non-planar phase reorganization,  hydrodynamic scaling is only gradually suppressed with $N\gtrsim7$, in contrast to the planar case where hydrodynamic scaling is suppressed for all $N\ge4$.

Our results provide several predictions to modulate coarsening dynamics which can be tested in experiments. For instance, recent advances in DNA nanotechnology have realized synthetic mixtures that exhibit up to nine coexisting liquid phases~\cite{chaderjian2025diversedistinctdenselypacked,Fabrini2024,GongAFM,Abraham_2024,BiffiPNAS}. Because they are assembled in dense monolayers, they naturally form planar contact networks, in accordance with our prediction of universal coarsening dynamics. Even in non-planar configurations, our results suggest that nine coexisting phases should be sufficient to significantly delay coalescence. These insights may provide design principles for wide-ranging applications based on this DNA nanotechnology, such as for droplet-based assays and model cell conjugates~\cite{Samanta2024,Chiara2016,Hirotake2023,D0CS00307G,Angelini}. In addition, recent works in biomolecular condensates highlight the possibility for multiple coexisting immiscible fluid phases in the cellular cytoplasm~\cite{FERIC20161686,Chen2024,Wan2018,doyle2023,SANDERS2020306,Ye2025.05.14.654140,Wan2018,Fare2021,Zhao2024,Fei2017}. An open question is to determine how cellular systems can select molecular interactions to generate functional target structures, given that these interactions determine both the number of phases and their interfacial tensions~\cite{zwicker,Rodrigo2024,teixeira2025metastablephaseseparationinformation}. Geometric confinement also offers a potential mechanism for cells to control the dominant coarsening regime to regulate spatial organization and design functional biomolecular condensate structures. 

There are also several key aspects of coarsening dynamics that we have not accounted for in this work. For example, future work should explore how surface wetting~\cite{Agudo-Canalejo2021,Mangiarotti2023} affects possible phase topologies in systems confined by solid structures. Many relevant complex fluids, especially biomolecular condensates, show viscoelastic and aging behaviors ~\cite{Mannattil,Tanaka2022, BERGERONSANDOVAL20184754}, which can influence the dominant coarsening mechanisms. More broadly, we anticipate that the contact network framework employed here can be beneficial for understanding systems with dynamically evolving domains, such as protein reaction-diffusion systems~\cite{Weyer2026}.

\vspace{0.5em}

\section*{Acknowledgements}
This work was supported by Leverhulme Trust (research project grant RPG-2022-140) and UKRI Engineering and Physical Sciences Research Council (EP/V034154/2). This work used the Cirrus UK National Tier-2 HPC Service at EPCC. We also acknowledge compute resources on ARCHER2 via the UK Consortium on Mesoscale Engineering Sciences (EP/L00030X/1).

\appendix

\section{Lattice Boltzmann Method}~\label{sec:LBE}

We use the phase field lattice Boltzmann method (LBM)~\cite{NCompMethod,lbmbook1} to solve the continuity, Navier-Stokes and Cahn-Hilliard equations. In the LBM, we consider distributions of fluid particles on lattice sites, free to move in a discrete number $Q$ of directions $\mathbf{e}_k$ according to a prescribed stencil. This stencil has associated weights $\omega_k$ in each direction. In this study, we use the D2Q9 stencil in 2D and the D3Q19 stencil in 3D~\cite{lbmbook1}. As is customary in the LBM, we take the lattice sound speed as $c_s^2=1/3$ and the time step $\delta t=1$. All quantities in this section are given in lattice units. For simplicity, we choose the BGK collision operator, although in practice other collision models can be used.

\subsection{Hydrodynamics}\label{sec:NmethH}
In this study, we assume that the total fluid mixture is incompressible. The Lattice Boltzmann Equation (LBE) to solve Eqs.~\ref{eqn:ce} and~\ref{eqn:nse} can be written as
\begin{align}
&f_{k}(\mathbf{x}+\mathbf{e}_k\delta_t,t+\delta_t)-f_{k}(\mathbf{x},t)=\notag\\&-\frac{1}{\tau_f}\left[f_{k}(\mathbf{x},t)-f^{eq}_{k}(\mathbf{x},t)\right]+\delta_t F_{k}(\mathbf{x},t).
\end{align}
This is used to update the fluid distribution function $f_k$, with a relaxation time of $\tau_f$. In this equation, the equilibrium distribution is given by
\begin{equation}
f^{eq}_{k}=\begin{cases}
    \frac{p_h}{c_s^2}(1-\omega_0)+\rho \left(s_k(\mathbf{v})\right),& \text{if } k=0,\\
    \frac{p_h}{c_s^2}\omega_i+\rho \left(s_k(\mathbf{v})\right),& \text{otherwise,}
\label{eq:feq}
\end{cases}
\end{equation}
where
\begin{equation}
s_k(\mathbf{v})=\omega_k\left[\frac{\mathbf{v}\cdot\mathbf{e}_k}{c_s^2}+\frac{\left(\mathbf{v}\cdot\mathbf{e}_k\right)^2}{2c_s^4}-\frac{\mathbf{v}\cdot\mathbf{v}}{2c_s^2}\right].
\end{equation}
The dynamic viscosity is given by
\begin{equation}
\eta=c_s^2\rho(\tau_f-0.5)\delta_t.
\end{equation}
The form of the force $F_{k}$ is given by~\cite{GuoForcing}
\begin{dmath}
\label{eqn:hydroforce}
F_{k}=\left(1-\frac{1}{2\tau_f}\right)\omega_i\left[\frac{\boldsymbol{v}\cdot\boldsymbol{\nabla}\rho c_s^2+\mathbf{e}_k\cdot \boldsymbol{F}_s}{c_s^2}+\frac{(\boldsymbol{v}\cdot\boldsymbol{\nabla}\rho c_s^2+\boldsymbol{v}\boldsymbol{F}_s):(\mathbf{e}_k\mathbf{e}_k-c_s^2\boldsymbol{I})}{c_s^4}\right],
\end{dmath}
where $:$ refers to the double dot product and $\boldsymbol{I}$ is the identity matrix. For this work, the body force corresponds to the thermodynamic force arising from the fluid mixing free energy 
\begin{equation}
\label{eq:Fsnew}
\boldsymbol{F}_s=\nabla{E_b}+\sum_{i,j=1;i\neq j}^N\kappa_{ij}\nabla^2C_j\nabla C_i.
\end{equation}
Thus, in our numerical algorithm, considering the pressure tensor in Eq.~\ref{eq:stress_tensor_compact}, the variation of $-\nabla p_h$ is applied through the equilibrium distribution, while the remainder of $-\nabla\mathbf{P}$ is implemented as an external force.

Unless otherwise specified, directional gradients $\nabla$ are given by a second order isotropic central difference scheme
\begin{equation}
\nabla C_i=\frac{1}{c_s^2\delta t}\sum_{k=1}^{Q-1}\omega_k\mathbf{e}_k[C_{i,x+\mathbf{e}_k\delta t}-C_{i,x}],
\end{equation}
and $\Delta C_i$ needed to evaluate $\mu_i$ is given by
\begin{equation}
\Delta C_i=\frac{2}{c_s^2\delta t^2}\sum_{k=1}^{Q-1}\omega_k[C_{i,x+\mathbf{e}_k\delta t}-C_{i,x}].
\end{equation}
Moments of the distribution function provide the momentum and hydrodynamic pressure
\begin{dmath}
    \rho\mathbf{v}=\sum_{k=0}^{Q-1}\mathbf{e}_kf_k+\frac{\Delta t}{2}\mathbf{F}_s,
\label{eq:nmom}
\end{dmath}
\begin{dmath}
    p_h=\frac{c_s^2}{1-\omega_k}\left[\sum_{k\neq0}^{Q-1}f_k+\frac{\delta t}{2}\mathbf{v}\cdot\boldsymbol{\nabla}\rho+\rho s_0(\mathbf{v})\right].
\label{eq:pres}
\end{dmath}

\subsection{Interface Capturing}\label{sec:NmethI}
The LBE to solve Eq.~\ref{eq:CHFlux} given by~\cite{wellbalanced1}
\begin{align}
&g_{i,k}(\mathbf{x}+\mathbf{e}_k\delta_t,t+\delta_t)-g_{i,k}(\mathbf{x},t)=\\&-\frac{1}{\tau_{g}}\left[g_{i,k}(\mathbf{x},t)-g^{eq}_{i,k}(\mathbf{x},t)\right]\notag\\&+\delta_t G_{i,k}(\mathbf{x},t)+\frac{1}{2}\delta_t^2 \partial_t G_{i,k}(\mathbf{x},t),
\label{eq:glbe}
\end{align}
with the equilibrium distribution $g^{eq}_{i,k}$ given by~\cite{wellbalanced1}
\begin{align}
&g^{eq}_{i,k}=\notag\\&\begin{cases}
    C_i+(1-\omega_0)m\sum_j^N\alpha_{ij}(\mu_j+\phi_j+\xi_j),& \text{if } k=0,\\
    -\omega_km\sum_j^N\alpha_{ij}(\mu_j+\phi_j+\xi_j),& \text{otherwise.}
\end{cases}
\end{align}
In this equation, $m$ is a parameter to control the mobility $M$. The advective term in the Cahn-Hilliard equation is implemented using the source term $G_{i,k}$~\cite{wellbalanced1}
\begin{equation}
G_{i,k}=\omega_k(\mathbf{v}\cdot\boldsymbol{\nabla}C_i)\left[-1+\frac{\boldsymbol{I}:(\mathbf{e}_k\mathbf{e}_k-c_s^2\boldsymbol{I})}{2c_s^2}\right].
\end{equation}
The time derivative in Eq.~\ref{eq:glbe} is calculated from a first order backwards difference scheme as $\partial_t G_{i,k}(\mathbf{x},t)=(G_{i,k}(\mathbf{x},t)-G_{i,k}(\mathbf{x},t-\delta t))/\delta t$. The fluid concentrations can be calculated from the zeroth moment of the distribution functions
\begin{equation}
C_i=\sum_{k=0}^{Q-1} g_{i,k}.
\end{equation}
The concentration $C_i$ can be used to calculate the fluid density and viscosity
\begin{equation}
\rho=\sum_{i=1}^N\rho_iC_i,
\end{equation}
\begin{equation}
\frac{1}{\tau_f}=\sum_{i=1}^N\frac{C_i}{\tau_i}.
\end{equation}
In this work we set $\rho_i=1$ and $\tau_i=1$ for all components. The mobility appearing in the Cahn-Hilliard equation is given by
\begin{equation}
M=c_s^2m(\tau_{g}-0.5)\delta_t.
\end{equation}
We vary $m$ to set the mobility and set $\tau_{g}=1$, with the exception of Fig.~\ref{fig:PhaseSep} (E,F) where we set $\tau_{g}=2.5$ to improve simulation stability during the initial demixing process.

\section{Procedure for Reduction Consistency}~\label{sec:reduct}

Ensuring reduction consistency - that absent fluids remain absent - is critical for stability and accuracy. We demonstrate this by simulating a three fluid separation process in an $N=4$ fluid model with $\phi_j=0$ in Eq.~\ref{eq:G}. Without reduction consistency, the absent fluid $C_4$ erroneously nucleates, shown in Fig. S4. Because we enforce mass conservation, the constraint $\int C_4dV=0$ leads to negative values of $C_4$ elsewhere to compensate for the positive $C_4$ nucleation, which quickly drives the system to instability.

To resolve this issue, we base our approach on the work by Boyer and Minjeaud (2014)~\cite{boyerncomp} and use the following form of the Cahn-Hilliard equation
\begin{equation}
\label{eq:CHGeneral}
\partial_t C_i+\nabla\cdot(\mathbf{v}C_i)= M\mathbf{\nabla}^2\left(\sum^N_{j=1}\alpha_{ij}(\mu_j+\phi_j+\xi_j) \right),
\end{equation}

where
\begin{align}
\mu_j = \sum_{k,k\neq j}^N\bigg[&\beta_{jk} \Big( f'(C_j) - f'(C_j + C_k) \Big)\notag\\ &+\kappa_{jk} \nabla^2 C_k\bigg],\label{eq:mu}\\f'(C_j) =& 2C_j(1-C_j)(1-2C_j)\notag  \end{align}
\begin{equation}
\phi_j=\frac{12}{D}\sum_{\substack{1\leq k<l<m\leq N\\k\neq j, l\neq j, m\neq j}}\Lambda_{j;{j,k,l,m}}C_kC_lC_m, \label{eq:G}
\end{equation}
\begin{equation}
\xi_j=\frac{12\Omega}{D}\sum_{\substack{1\leq k<l\leq N\\k\neq j, l\neq j}}C_k^2C_l^2\bigg(2C_j-\sum_{\substack{1\leq m\leq N\\m\neq j, k, l}}\Theta_{kl;j;m}C_m\bigg), \label{eq:P} 
\end{equation}
where $\Omega$ is a free parameter chosen to improve simulation stability with large interfacial tension contrasts. The coefficients $\alpha_{ij}$ are given by solving the linear system
\begin{equation}
\boldsymbol{\alpha\sigma}=\boldsymbol{I}+\boldsymbol{\gamma}\otimes\boldsymbol{1}^T,
\label{eq:alpha1}
\end{equation}
\begin{equation}
\boldsymbol{\alpha}\boldsymbol{1}=\boldsymbol{0},
\label{eq:alpha2}
\end{equation}
where $\boldsymbol{\gamma}$ is a vector of size $N$ to be determined and $\boldsymbol{\sigma}$ is the matrix of interfacial tensions. Additionally, $\otimes$ is the Kronecker product and $\boldsymbol{1}$ and $\boldsymbol{0}$ refer to a vector of size $N$ with each entry as $1$ and $0$ respectively.

To obtain reduction consistency, we must choose $\Lambda_{j;{j,k,l,m}}$ to satisfy
\begin{dmath}
\sum_{\substack{1\leq j\leq N\\j\neq k,l,m}}\alpha_{ij}\Lambda_{j;{j,k,l,m}}=\Gamma_{i,k,l,m},\ \forall i\hiderel\notin{k,l,m},
\end{dmath}
\begin{dmath}
\label{eq:ngamma}
\Gamma_{i,k,l,m}=6\left(\alpha_{ik}(\sigma_{kl}+\sigma_{km})+\alpha_{il}(\sigma_{kl}+\sigma_{lm})\hiderel+\alpha_{im}(\sigma_{km}+\sigma_{lm})-\gamma_i\right).
\end{dmath}
The result is a linear system that can be solved to calculate $\Lambda_{j;{j,k,l,m}}$ following
\begin{equation}
(\boldsymbol{\alpha}_{i,j})(\boldsymbol{\Lambda}_j)=(\boldsymbol{\Gamma}_i),\ \forall i,j\hiderel\notin{k,l,m}.
\end{equation}
$(\boldsymbol{\alpha}_{i.j})$ refers to the submatrix of $\alpha$ formed of all entries $i,j$ not equal to $k,l,m$. $(\boldsymbol{\Lambda}_j)$ and $(\boldsymbol{\Gamma}_i)$ are vectors of size $N-3$ with each entry as $\Lambda_{j;{j,k,l,m}}$ and $\Gamma_i$ respectively for all $i,j$ not equal to $k,l,m$. We have one of such linear systems for each set of $k,l,m$, which can together be used to find all $\Lambda_{j;{j,k,l,m}}$.

The inclusion of the stabilising contribution $\xi_j$~\cite{boyerncomp} requires the calculation of $\Theta_{kl;j;m}$, which are obtained from
\begin{dmath}
\sum_{\substack{1\leq j\leq N\\j\neq k,l,m}}\alpha_{ij}\Theta_{kl;j;m}=2\alpha_{im},\ \forall i\hiderel\notin{k,l,m}.
\end{dmath}
Here, we can solve the linear system
\begin{equation}
(\boldsymbol{\alpha}_{i,j})(\boldsymbol{\Theta}_{k,l})=2(\boldsymbol{\alpha}_{i,m}),\ \forall i,j\hiderel\notin{k,l,m},
\end{equation}
to obtain all $\Theta_{kl;j;m}$ for a given $k,l,m$.

\section{Construction of Contact Networks}~\label{sec:contactnetwork}

We identify contact networks using the following algorithm. Step one: for each lattice point, we label the immiscible fluid component with the largest concentration. Step two: starting on a given lattice point, we label it with a unique id and take note of its component label. Then, we look at the neighbors of that lattice point. If they match the stored component, we label them with the same id and add them to a queue. For each lattice point in the queue, we then repeat this process, deleting lattice points in the queue after we visit them, until we have run out of lattice points in the queue. At this point, we repeat step two starting with a lattice point that is not already in any of the identified connected region. Step three: we extract a list of edges by storing the set of every two regions that share at least one pair of neighbouring lattice points. Taken together, we can construct the contact network.

From the contact network, we can also identify the building blocks in Fig.~\ref{fig:2D} (A) as follows. Starting on any given node, we store that node id and the id of all nodes that are directly connected to it by an edge. Then, we identify the building block as the graph containing these nodes and all edges that connect two of the stored nodes. We repeat this for each node to find all unique building blocks.

\section{Coalescence Probability Measurement}~\label{sec:coalesce}
To identify coalescence and Ostwald ripening events, we first extract contact network using the approach in Appendix C at a given timestep. We then calculate centre of mass and area of each region associated with each node in the network. Snapshots of the simulations are recorded every $\Delta t = 0.074$. To minimise memory requirements due to frequent saving, we use a periodic unit size of $165\times165$. Between each consecutive snapshot, we match each region based on distances between the centre of masses in subsequent snapshots. The regions without a match within a distance of $0.667$ are considered to have disappeared or coalesced, with $0.667$ chosen to be larger than the maximum velocity in all simulations over $\Delta t = 0.074$. We differentiate Ostwald ripening and coalescence based on the area of the region in the previous frames. Regions with an area of less than $\approx0.01\%$ of the simulation domain area are considered to have disappeared due to Ostwald ripening. We count one coalescence event for every two domains that disappear due to coalescence.

We evaluate the coalescence probability by counting the number of coalescence events divided by the current number of fluid regions over time and summing these events over length scale intervals of $\Delta L^*=0.5$. The results are then rescaled by the time taken for the length scale to increase from $L$ to $L+0.5$ for each $N$ to obtain the coalescence probability over an interval of $\Delta t=1$.

\section{Evaluation of $a_N$}~\label{sec:aN}

Following Marqusee~\cite{marqusee}, the steady state (scaled) droplet size distribution is given by
\begin{align}
F_0(z)\propto\frac{1}{\omega(z)}\exp\!\!\left[\!\int_0^z\!\frac{dz'}{\omega(z')}\!\right],\\
\int_0^\infty z^2F_0(z)\,dz=1,\quad F_0(z>z_0)=0, \nonumber
\end{align}
where $z=R/\langle R\rangle$ is the scaled droplet radius and the growth rate is 

\begin{equation}
\omega(z)=\frac{3}{2s_0}\frac{K_1(z/s_0)}{K_0(z/s_0)}\!\left(\sigma_1-\frac{1}{z}\right)-\frac{z}{2}.
\end{equation}
$K_0$ and $K_1$ are the modified Bessel functions of the second kind. $\sigma_1$ can be interpreted as the amplitude of the decay in supersaturation, and together with the limit $z_0$, they are fixed by the conditions $\omega(z_0)=0$ and $\partial_z\omega(z_0)=0$. Then, the screening length $s_0$, which captures the effective finite-range interactions between the domains, can be obtained by iteratively solving for 
\begin{equation}
s_0^{-1}=2V_N\!\int_0^\infty\!\frac{zK_1(z/s_0)}{K_0(z/s_0)}F_0(z)\,dz.
\end{equation}
$V_N$ is the volume fraction, which is equal to $V_N=1/N$ for the equal composition scenario. 
The prefactor can then be computed as $a_N=\int zF_0/\int F_0$. This allows us to obtain $a_3=0.975$, $a_4=0.923$, $a_5=0.887$ and $a_6=0.859$.

\section{Estimation of $k_{\mathrm{e}}$}~\label{sec:ke}

To estimate $k_{\mathrm{e}}$ in Eq.~\eqref{eq:ke}, we directly measure the rates of coalescence and Ostwald ripening between $t=185$ to $t=333$ for $N\geq3$, using the approach in Appendix~\ref{sec:coalesce}. We again use a periodic unit size of $165\times165$, and we measure coalescence events over intervals of $\Delta t = 0.148$.
All other parameters are identical to those in Section~\ref{sec:planar}. We then count the number of coalescence events per Ostwald ripening event $\langle n_{\mathrm{c}}\rangle$ over all timesteps, and fit the numerical results against Eq.~\ref{eq:ncprop} to get $k_{\mathrm{e}}=0.402$.

\bibliography{apssamp}

\end{document}